\documentclass[11pt]{article}
\usepackage{amsmath}
\usepackage{amscd}
\usepackage{graphicx}
\usepackage{latexsym}
\usepackage{amsmath,amsthm,amsfonts,amssymb,cite}%,showlabels}
\usepackage{latexsym}
\usepackage{amsmath}

\theoremstyle{remark}

\numberwithin{equation}{section}
\begin{document}
\hoffset = -2.4truecm \voffset = -2truecm
\renewcommand{\baselinestretch}{1.2}
\newcommand{\mb}{\makebox[10cm]{}\\ }
\date{}

%%%%% DOCUMENT SPECIFIC DEFINITIONS

%  Theorems, Lemmas and the like, should be typeset in italic

\newtheorem{theorem}{Theorem}

\newtheorem{proposition}{Proposition}

\newtheorem{lemma}{Lemma}

\newtheorem{definition}{Definition}

%%%%% END DOCUMENT SPECIFIC DEFINITIONS

%\renewcommand{\square}{\hfill$\Box$\vspace{2ex}}

%\renewcommand{\Theta}{\Ta}

\title{A noncommutative semi-discrete Toda equation and its quasideterminant solutions}

\author{C.X. Li$^{1}$ and J.J.C. Nimmo$^{2}$\\
$^{1}$School of Mathematical Sciences,\\
Capital Normal University\\
Beijing 100037, CHINA\\
$^{2}$Department of Mathematics, \\
University of Glasgow \\
Glasgow G12 8QW, UK }

\date{}

 \maketitle

\begin{abstract}
A noncommutative version of the semi-discrete Toda equation is
considered. A Lax pair and its Darboux transformations and binary
Darboux transformations are found and they are used to construct two
families of quasideterminant solutions.
\end{abstract}

\section{Introduction}
Recently, there has been a lot of interest in noncommutative
versions of some well-known soliton equations, such as the KP
equation, the KdV equation, the Hirota-Miwa equation, the modified
KP equation and the two-dimensional Toda lattice
\cite{K,P,S,WW1,WW2,WW3,H,HT1,DH,JN,GN1,GN2,GNS,LN}. There are a
number of reasons for this lack of commutativity. For example, the
variables might be square matrices or quarternions and so on.
Another natural way in which the variables fail to commute is
because of a quantization of the phase space resulting in the normal
product being replaced by a Moyal product. In the approach taken
here, it is not necessary to specify the reason for the lack of
commutivity. Often, the noncommutative version is obtained simply by
assuming that the coefficients in the Lax pair of the commutative
equation do not commute.

Quasideterminants \cite{GR,EGR,GGRL} play the much same role in
noncommutative algebra as determinants do in standard, commutative
algebra. They arise in many situations where determinants appear in
commutative algebra and this suggests that they are a very natural
structure to use when working with noncommutative integrable
systems. In particular, determinants are ubiquitous as solutions of
commutative integrable systems and one way to obtain these is
through the use of Darboux or binary Darboux transformations.
Quasideterminant solutions of noncommutative versions of these
integrable systems arise in the same way. The reader is referred to
the original papers \cite{GR,EGR,GGRL} for a detailed and general
treatment of quasideterminants. See also \cite{LN} for a summary of
the key results used in the current paper.

The semi-discrete Toda equation
\begin{equation}\label{ST}
\dfrac{d}{dt}\log\left({v_n^{k+1}\over
v_n^k}\right)=v_{n+1}^{k+1}+v_{n-1}^k-v_n^k-v_n^{k+1},
\end{equation}
was first considered in \cite{RI}. Some other related results were
presented in \cite{LH,ZGH,WHY}.

If one writes
\[
    v_n^k={\tau_{n+1}^{k+1}\tau_{n-1}^k\over\tau_n^{k+1}\tau_n^k},
\]
and $\tau_n^k$ satisfies the Hirota bilinear equation
\begin{equation}\label{BST}
D_t\tau_n^{k+1}\cdot\tau_n^{k}-\tau_{n+1}^{k+1}\tau_{n-1}^{k}+\tau_n^{k+1}\tau_n^{k}=0,
\end{equation}
then $v_n^k$ satisfies \eqref{ST}. In \eqref{BST} $D_t$ denotes Hirota's bilinear operator defined by \cite{HR2}
\[
D_t^na(t)\cdot b(t)\equiv \left (\frac {\partial }{\partial t}-\frac
{\partial }{\partial t'}\right )^na(t)b(t')|_{t'=t}.
\]

This paper is concerned with the following noncommutative generalization of \eqref{ST},
\begin{align}
\label{dt1}
&v_{n}^ku_{n}^k=u_{n}^{k+1}v_{n+1}^k,\\
\label{dt2}
&v_{n+1,t}^k+u_{n+1}^{k+1}-u_n^k=0,
\end{align}
in which $u_n$ and $v_n$ and their derivatives do not commute in general. By introducing new variable $X_n^k$ where
$$u_n^k=X_n^k(X_{n+1}^k)^{-1},\quad v_n^k=X_n^{k+1}(X_n^k)^{-1}$$
\eqref{dt1} is satisfied identically and \eqref{dt2} becomes
\begin{equation}
(X_{n+1}^{k+1}(X_{n+1}^k)^{-1})_t+X_{n+1}^{k+1}(X_{n+2}^{k+1})^{-1}-X_n^{k}(X_{n+1}^k)^{-1}=0.\label{NCST}
\end{equation}
In the commutative reduction, it is easy to show, by writing
$X_n^k=\tau_{n-1}^k/\tau_n^k$, that \eqref{NCST} becomes
\begin{eqnarray*}
&&\tau_{n+1}^{k+1}\tau_{n+1}^k(D_t\tau_n^{k+1}\cdot\tau_n^k-\tau_{n+1}^{k+1}\tau_{n-1}^k+\tau_n^{k+1}\tau_n^k)\\
&&
-\tau_n^k\tau_n^{k+1}(D_t\tau_{n+1}^{k+1}\cdot\tau_{n+1}^k-\tau_{n+2}^{k+1}\tau_{n}^k+\tau_{n+1}^{k+1}\tau_{n+1}^k)=0,
\end{eqnarray*}
which is satisfied whenever the bilinear semi-discrete Toda equation
\eqref{BST} is satisfied, thus verifying that indeed
\eqref{dt1}--\eqref{dt2} is a noncommutative generalization of
\eqref{ST}.  For the rest of this paper, we will refer to
\eqref{NCST} as the noncommutative semi-discrete Toda equation.

The main results of this paper are to show that this noncommutative
system is integrable in the sense that it has a Lax pair and the
associated Darboux and binary Darboux transformations may be
iterated to construct families of exact solutions. We show how these
solutions may be expressed in terms of quasideterminants of two
different types.

The paper is organized as follows. In Section~2, we present a Lax
pair and its Darboux transformation and describe how iteration of
this transformation gives quasicasoratian solutions. An adjoint
linear problem and binary Darboux transformations are discussed in
Section~3, and quasigrammian solutions are obtained. Conclusions are
given in Section~4.

\section{Quasicasoratian solutions obtained by Darboux transformations}
The non-commutative semi-discrete Toda lattice \eqref{NCST} has Lax
pair
\begin{align}
\phi_{n,t}^k&=X_n^k(X_{n+1}^k)^{-1}\phi_{n+1}^k,\label{LP1}\\
\phi_{n+1}^{k+1}&=\phi_n^k-X_{n+1}^{k+1}(X_{n+1}^k)^{-1}\phi_{n+1}^k.\label{LP2}
\end{align}
This was obtained by discretizing the Lax pair of the non-Abelian
Toda lattice \cite{MK,NW} and simplifying the Lax pair of the
semi-discrete Toda equation with self-consistent sources \cite{WHY}.

Let $\theta_{n,i}^k, i=1,\ldots,N$ be a particular set of
eigenfunctions of the linear system and from these define the row
vector $\Theta_n^k=(\theta_{n,1}^k,\ldots,\theta_{n,N}^k)$. The
Darboux transformation, determined by the particular solution
$\theta_n^k$, for the noncommutative semi-discrete Toda lattice is
\begin{align*}
\widetilde{\phi}_n^k&=\phi_n^k-\theta_n^k(\theta_{n+1}^k)^{-1}\phi_{n+1}^k,\\
%\tilde{u}_n^k&=u_n^k-(\theta_n^k(\theta_{n+1}^k)^{-1})_t=\theta_n^k(\theta_{n+1}^k)^{-1}u_{n+1}^k\theta_{n+2}^k(\theta_{n+1}^k)^{-1},\\
%\tilde{v}_n^k&=v_n^k+\theta_n^{k+1}(\theta_{n+1}^{k+1})^{-1}-\theta_{n-1}^k(\theta_n^k)^{-1}=\theta_n^{k+1}(\theta_{n+1}^{k+1})^{-1}v_{n+1}^k\theta_{n+1}^k(\theta_n^k)^{-1}
%=\theta_n^{k+1}(\theta_n^k)^{-1}v_{n+1}^k\theta_{n+1}^k(\theta_{n+1}^{k+1})^{-1},\\
\widetilde{X}_n^k&=\theta_n^k(\theta_{n+1}^k)^{-1}X_{n+1}^k.
\end{align*}
This transformation may be iterated by defining
\begin{align}
\phi_n^k[l+1]&=\phi_n^k[l]-\theta_n^k[l](\theta_{n+1}^k[l])^{-1}\phi_{n+1}^k[l],\label{DT1}\\
X_n^k[l+1]&=\theta_{n}^k[l](\theta_{n+1}^k[l])^{-1}X_{n+1}^k[l],\label{DT2}
\end{align}
where $\phi_n^k[1]=\phi_n^k, X_n^k[1]=X_n^k$ and
\begin{equation}
\theta_{n}^k[l]=\phi_n^k[l]|_{\phi_n^k\rightarrow \theta_{n,l}^k}.
\end{equation}
In particular,
\begin{align}
\phi_n^k[2]&=\phi_n^k-\theta_{n,1}^k(\theta_{n+1,1}^k)^{-1}\phi_{n+1}^k,\label{DT10}\\
X_n^k[2]&=\theta_{n,1}^k(\theta_{n+1,1}^k)^{-1}X_{n+1}^k.\label{DT20}
\end{align}

In what follows, we shall show by induction that the results of $N$
repeated Darboux transformations, $\phi_n^k[N+1]$ and $X_n^k[N+1]$,
can be expressed in closed form as quasideterminants
\begin{equation}
\phi_n^k[N+1]=
\begin{vmatrix}
\Theta_{n}^k&\fbox{$\phi_n^k$}\\
\Theta_{n+1}^k&\phi_{n+1}^k\\
\vdots&\vdots\\
\Theta_{n+N}^k&\phi_{n+N}^k
\end{vmatrix},\quad
X_n^k[N+1]=(-1)^{N}
\begin{vmatrix}
\Theta_n^k&\fbox{$0$}\\
\Theta_{n+1}^k&0\\
\vdots&\vdots\\
\Theta_{n+N}^k&1
\end{vmatrix}X_{n+N}^k.\label{QW}
\end{equation}

The initial case $N=1$ follows directly from
\eqref{DT10}--\eqref{DT20}. Also, using the noncommutative Jacobi
identity, row homological relations and definition of
quasi-Pl\"ucker coordinates (see (2.3)--(2.5) in \cite{LN} for
example), we have
\begin{align*}
\phi_n^k[N+2]&=\phi_n^k[N+1]-\theta_n^k[N+1]\theta_{n+1}^k[N+1]^{-1}\phi_{n+1}^k[N+1]\\
&=\begin{vmatrix}
\Theta_n^k&\fbox{$\phi_n^k$}\\
\Theta_{n+1}^k&\phi_{n+1}^k\\
\vdots&\vdots\\
\Theta_{n+N}^k&\phi_{n+N}^k
\end{vmatrix}-\begin{vmatrix}
\Theta_{n}^k&\fbox{$\theta_{n,N+1}^k$}\\
\Theta_{n+1}^k&\theta_{n+1,N+1}^k\\
\vdots&\vdots\\
\Theta_{n+N}^k&\theta_{n+N,N+1}^k
\end{vmatrix}\begin{vmatrix}
\Theta_{n+1}^k&\fbox{$\theta_{n+1,N+1}^k$}\\
\Theta_{n+2}^k&\theta_{n+2,N+1}^k\\
\vdots&\vdots\\
\Theta_{n+N+1}^k&\theta_{n+N+1,N+1}^k
\end{vmatrix}^{-1}\begin{vmatrix}
\Theta_{n+1}^k&\fbox{$\phi_{n+1}^k$}\\
\Theta_{n+2}^k&\phi_{n+2}^k\\
\vdots&\vdots\\
\Theta_{n+N+1}^k&\phi_{n+N+1}^k
\end{vmatrix}\\
&=\begin{vmatrix}
\Theta_{n}^k&\fbox{$\phi_n^k$}\\
\Theta_{n+1}^k&\phi_{n+1}^k\\
\vdots&\vdots\\
\Theta_{n+N}^k&\phi_{n+N}^k
\end{vmatrix}-\begin{vmatrix}
\Theta_{n}^k&\fbox{$\theta_{n,N+1}^k$}\\
\Theta_{n+1}^k&\theta_{n+1,N+1}^k\\
\vdots&\vdots\\
\Theta_{n+N}^k&\theta_{n+N,N+1}^k
\end{vmatrix}
\begin{vmatrix}
\Theta_{n+1}^k&\theta_{n+1,N+1}^k\\
\Theta_{n+2}^k&\theta_{n+2,N+1}^k\\
\vdots&\vdots\\
\Theta_{n+N+1}^k&\fbox{$\theta_{n+N+1,N+1}^k$}
\end{vmatrix}^{-1}\begin{vmatrix}
\Theta_{n+1}^k&\phi_{n+1}^k\\
\Theta_{n+2}^k&\phi_{n+2}^k\\
\vdots&\vdots\\
\Theta_{n+N+1}^k&\fbox{$\phi_{n+N+1}^k$}
\end{vmatrix}\\
&=\begin{vmatrix}
\Theta_{n}^k&\theta_{n,N+1}^k&\fbox{$\phi_n^k$}\\
\Theta_{n+1}^k&\theta_{n+1,N+1}^k&\phi_{n+1}^k\\
\vdots&\vdots\\
\Theta_{n+N+1}^k&\theta_{n+N+1,N+1}^k&\phi_{n+N+1}^k
\end{vmatrix}.
\end{align*}
In a similar way, we also have
\begin{align*}
X_n^k[N+2]&=\theta_{n}^k[N+1]\theta_{n+1}^k[N+1]^{-1}X_{n+1}^k[N+1]\\
&=(-1)^N\begin{vmatrix}
\Theta_{n}^k&\fbox{$\theta_{n,N+1}^k$}\\
\Theta_{n+1}^k&\theta_{n+1,N+1}^k\\
\vdots&\vdots\\
\Theta_{n+N}^k&\theta_{n+N,N+1}^k
\end{vmatrix}\begin{vmatrix}
\Theta_{n+1}^k&\fbox{$\theta_{n+1,N+1}^k$}\\
\Theta_{n+2}^k&\theta_{n+2,N+1}^k\\
\vdots&\vdots\\
\Theta_{n+N+1}^k&\theta_{n+N+1,N+1}^k
\end{vmatrix}^{-1}\begin{vmatrix}
\Theta_{n+1}^k&\fbox{$0$}\\
\Theta_{n+2}^k&0\\
\vdots&\vdots\\
\Theta_{n+N+1}^k&1
\end{vmatrix}X_{n+N+1}^k\\
&=(-1)^N\begin{vmatrix}
\Theta_{n}^k&\fbox{$\theta_{n,N+1}^k$}\\
\Theta_{n+1}^k&\theta_{n+1,N+1}^k\\
\vdots&\vdots\\
\Theta_{n+N}^k&\theta_{n+N,N+1}^k
\end{vmatrix}\begin{vmatrix}
\Theta_{n+1}^k&\theta_{n+1,N+1}^k\\
\vdots&\vdots\\
\Theta_{n+N}^k&\theta_{n+N,N+1}^k\\
\Theta_{n+N+1}^k&\fbox{$\theta_{n+N+1,N+1}^k$}
\end{vmatrix}^{-1}X_{n+N+1}^k\\
%\intertext{and then using the quasi-Pl\"ucker coordinate formula }
&=(-1)^{N+1}\begin{vmatrix}
\Theta_{n}^k&\theta_{n,N+1}^k&\fbox{$0$}\\
\Theta_{n+1}^k&\theta_{n+1,N+1}^k&0\\
\vdots&\vdots\\
\Theta_{n+N+1}^k&\theta_{n+N+1,N+1}^k&1
\end{vmatrix}X_{n+N+1}^k.
\end{align*}
This proves the inductive step and the proof is complete.

\section{Quasigrammian solutions obtained by binary Darboux transformations}
The linear equations \eqref{LP1} and \eqref{LP2} have the formal
adjoint
\begin{align}
-\psi_{n,t}^k&={u_{n-1}^k}^\dagger\psi_{n-1}^k,\label{ALP1}\\
\psi_{n+1}^{k+1}&=\psi_n^k+{v_n^k}^\dagger\psi_n^{k+1}.\label{ALP2}
\end{align}
As always with discrete Lax equations, some care is needed in
defining their adjoint. The key point is that they should be chosen
in such a way that the compatibility condition for
\eqref{ALP1}--\eqref{ALP2} is identical to that of
\eqref{LP1}--\eqref{LP2}. In other words, \eqref{ALP1}--\eqref{ALP2}
also form a Lax pair for \eqref{ST}.

Following the standard construction of a binary Darboux
transformation, one introduces a potential
$\Omega_n^k=\Omega(\phi_n^k,\psi_n^k)$ satisfying the three
conditions
\begin{align}
&\Omega_{n,t}^k=-{\psi_n^k}^\dagger u_n^k\phi_{n+1}^k,\label{NP1}\\
&\Omega_{n}^{k+1}-\Omega_{n}^k={\psi_{n+1}^{k+1}}^\dagger\phi_n^k,\label{NP2}\\
&\Omega_{n+1}^k-\Omega_{n}^k=-{\psi_{n+1}^k}^\dagger\phi_{n+1}^k.\label{NP3}
\end{align}

A binary Darboux transformation is then defined by
\begin{align}
\phi_n^k[N+1]&=\phi_n^k[N]-\theta_n^k[N]\Omega(\theta_n^k[N],\rho_n^k[N])^{-1}\Omega(\phi_n^k[N],\rho_n^k[N]),\label{BDT1}\\
\psi_n^k[N+1]&=\psi_n^k[N]-\rho_n^k[N]\Omega(\theta_{n-1}^k[N],\rho_{n-1}^k[N])^{-\dagger}\Omega(\theta_{n-1}^k[N],\psi_{n-1}^k[N])^\dagger,\label{BDT2}\\
X_n[N+1]&=(I+\theta_n^k[N]\Omega(\theta_n^k[N],\rho_n^k[N])^{-1}\rho_n^k[N]^\dagger)X_n^k[N],\label{BDT3}
\end{align}
where $\phi_n^k[1]=\phi_n^k,\ \psi_n^k[1]=\psi_n^k$,\
$X_n^k[1]=X_n^k$ and
\begin{equation}
\theta_n^k[N]=\phi_n^k[N]|_{\phi_n^k\rightarrow\theta_{n,N}^k},
\quad \rho_n^k[N]=\psi_n^k[N]|_{\psi_n^k\rightarrow\rho_{n,N}^k}.
\end{equation}

Using the notation
$\Theta_n^k=(\theta_{n,1}^k,\dots,\theta_{n,N}^k)$ and
$P_n^k=(\rho_{n,1}^k,\dots,\rho_{n,N}^k)$, it is easy to prove by
induction that for $N\ge 1$,
\begin{align}
\phi_n^k[N+1]&=\begin{vmatrix}
\Omega(\Theta_n^k,P_n^k)&\Omega(\phi_n^k,P_n^k)\\
\Theta_n^k&\fbox{$\phi_n^k$} \end{vmatrix},\label{NE}\\
\psi_n^k[N+1]&=\begin{vmatrix}
\Omega(\Theta_{n-1}^k,P_{n-1}^k)^\dagger&\Omega(\Theta_{n-1}^k,\psi_{n-1}^k)^\dagger\\
P_n^k&\fbox{$\psi_n^k$}
\end{vmatrix}\label{NAE}
\end{align}
and
\begin{equation} \Omega(\phi_n^k[N+1],\psi_n^k[N+1])=
\begin{vmatrix}
\Omega(\Theta_n^k,P_n^k)&\Omega(\phi_n^k,P_n^k)\\
\Omega(\Theta_n^k,\psi_n^k)&\fbox{$\Omega(\phi_n^k,\psi_n^k)$}
\end{vmatrix}.\label{NP}
\end{equation}
We may thus after $N$ binary Darboux transformations obtain
\begin{equation}
X_n^k[N+1]=-\begin{vmatrix}
\Omega(\Theta_n^k,P_n^k)&{P_n^k}^\dagger\\
\Theta_n^k&\fbox{$-I$} \end{vmatrix}X_n^k.\label{QG}
\end{equation}

In fact, we can prove the above results by induction. Using
\eqref{NE}--\eqref{NP}, we have
\begin{align*}
X_n^k[N+2]&=(I+\Theta_n^k[N+1]\Omega(\Theta_n^k[N+1],P_n^k[N+1])^{-1}P_n^k[N+1]^\dagger)X_n^k[N+1]\\
&=-\left(I+\begin{vmatrix}
\Omega(\Theta_n^k,P_n^k)&\Omega(\theta_{n,N+1}^k,P_n^k)\\
\Theta_n^k&\fbox{$\theta_{n,N+1}^k$} \end{vmatrix}
\begin{vmatrix}
\Omega(\Theta_n^k,P_n^k)&\Omega(\theta_{n,N+1}^k,P_n^k)\\
\Omega(\Theta_n^k,\rho_{n,N+1}^k)&\fbox{$\Omega(\theta_{n,N+1}^k,\rho_{n,N+1}^k)$}
\end{vmatrix}^{-1}\right.\\
&\quad\left.\begin{vmatrix}
\Omega(\Theta_{n-1}^k,P_{n-1}^k)&{P_n^k}^\dagger\\
\Omega(\Theta_{n-1}^k,\rho_{n-1,N+1}^k)&\fbox{${\rho_{n,N+1}^k}^\dagger$}
\end{vmatrix}\right)
\begin{vmatrix}
\Omega(\Theta_n^k,P_n^k)&{P_n^k}^\dagger\\
\Theta_n^k&\fbox{$-I$}
\end{vmatrix}X_n^k.
\end{align*}
Noticing that
\begin{align*}
&\begin{vmatrix}
\Omega(\Theta_{n-1}^k,P_{n-1}^k)&{P_n^k}^\dagger\\
\Omega(\Theta_{n-1}^k,\rho_{n-1,N+1}^k)&\fbox{${\rho_{n,N+1}^k}^\dagger$}
\end{vmatrix}
\begin{vmatrix}
\Omega(\Theta_n^k,P_n^k)&{P_n^k}^\dagger\\
\Theta_n&\fbox{$-I$}
\end{vmatrix}\\
&=-({\rho_{n,N+1}^k}^\dagger-\Omega(\Theta_{n-1}^k,\rho_{n-1,N+1}^k)\Omega(\Theta_{n-1}^k,P_{n-1}^k)^{-1}{P_n^k}^\dagger)
(I+\Theta_n^k\Omega(\Theta_n^k,P_n^k)^{-1}{P_n^k}^\dagger)\\
&=-{\rho_{n,N+1}^k}^\dagger+\Omega(\Theta_{n-1}^k,\rho_{n-1,N+1}^k)\Omega(\Theta_{n-1}^k,P_{n-1}^k)^{-1}{P_n^k}^\dagger\\
&\quad+(\Omega(\Theta_{n}^k,\rho_{n,N+1}^k)-\Omega(\Theta_{n-1}^k,\rho_{n-1,N+1}^k))\Omega(\Theta_n^k,P_n^k)^{-1}{P_n^k}^\dagger\\
&\quad+\Omega(\Theta_{n-1}^k,\rho_{n-1,N+1}^k)\Omega(\Theta_{n-1}^k,P_{n-1}^k)^{-1}
(\Omega(\Theta_{n-1}^k,P_{n-1}^k)-\Omega(\Theta_n^k,P_n^k))\Omega(\Theta_n^k,P_n^k)^{-1}{P_n^k}^\dagger\\
&=-{\rho_{n,N+1}^k}^\dagger+\Omega(\Theta_{n}^k,\rho_{n,N+1}^k)\Omega(\Theta_n^k,P_n^k)^{-1}{P_n^k}^\dagger\\
&=-\begin{vmatrix}
\Omega(\Theta_n^k,P_n^k)&{P_n^k}^\dagger\\
\Omega(\Theta_{n}^k,\rho_{n,N+1}^k)&\fbox{${\rho_{n,N+1}^k}^\dagger$}
\end{vmatrix},
\end{align*}
it follows that
\begin{equation*}
X_n^k[N+2]=-\begin{vmatrix}
\Omega(\Theta_n^k,P_n^k)&\Omega(\theta_{n,N+1}^k,P_n^k)&{P_n^k}^\dagger\\
\Omega(\Theta_n^k,\rho_{n,N+1}^k)&\Omega(\theta_{n,N+1}^k,\rho_{n,N+1}^k)&{\rho_{n,N+1}^k}^\dagger\\
\Theta_n^k&\theta_{n,N+1}^k&\fbox{$-I$}
\end{vmatrix}X_n^k,
\end{equation*}
as required.

\section{Conclusions}
In this paper, we have described a noncommutative version of the
semi-discrete Toda equation. We have obtained quasi\-casoratian and
quasigrammian solutions by means of discrete Darboux transformations
and binary Darboux transformations, respectively. We have given an
inductive proof of the iterated Darboux transformations and
solutions by using quasideterminant identities. Since we have not at
any point specified the nature of noncommutativity, the results
obtained here are valid whatever the reason for noncommutativity is.


\begin{thebibliography}{99}
\bibitem{K}B.A. Kupershmidt, KP or mKP: Noncommutative Mathematics
of Lagrangian, Hamiltonian, and Integrable systems. Mathematical
Surverys and Monographs (American Mathematical Society, New York),
78, 2000.
\bibitem{P}L.D. Paniak, Exact noncommutative KP and KdV multi-solitons, arXiv:hep-th/0105185
(2001).
\bibitem{S}M. Sakakibara, J. Phys. A 37 (2004) L599-L604.
\bibitem{WW1}N. Wang, M. Wadati, J. Phys. Soc. Jpn 72 (2003)
1366-1373.
\bibitem{WW2}N. Wang, M. Wadati, J. Phys. Soc. Jpn 72 (2003)
1881-1888.
\bibitem{WW3}N. Wang, M. Wadati, J. Phys. Soc. Jpn 73 (2004)
1689-1698.
\bibitem{H} M. Hamanaka, Noncommutative solitons and D-branes. PhD
thesis. arXiv:hep-th/0303256 (2003).
\bibitem{HT1}M. Hamanaka, Nucl. Phys. B 316 (2003) 77-83.
\bibitem{DH}A. Dimakis, M\"uller-Hoissen, J. Phys. A 38 (2005)
5453-5505.
\bibitem{JN}J.J.C. Nimmo, J. Phys. A 39 (2006) 5053-5065.
\bibitem{GN1}C.R. Gilson, J.J.C. Nimmo, J. Phys. A 40 (2007)
3839-3850.
\bibitem{GN2}C.R. Gilson, J.J.C. Nimmo, Y. Ohta, J. Phys. A: Math.
Theor. 40 (2007) 12607-12617.
\bibitem{GNS}C.R. Gilson, J.J.C. Nimmo, C.M. Sooman, J. Phys. A:
Math. Theor. 41 (2008) 085202.
\bibitem{LN}C.X. Li, J.J.C. Nimmo, Proc. R. Soc. A 464 (2008)
951-966.
\bibitem{GR}I.M. Gelfand, V.S. Retakh, Funkt. Anal. Prilozhen. 25
(1991) 13-25.
\bibitem{EGR}P. Etingof, I.M. Gelfand, V.S. Retakh, Math. Res.
Lett. 5(1998) 1-12.
\bibitem{GGRL}I.M. Gelfand, S. Gelfand, V.M. Retakh, R.L. Wilson,
Adv. Math. 193 (2005) 56-141.
\bibitem{RI}R. Inoue, K. Hikami, J. Phys. A: Math. Gen. 32 (1999)
6853-6868.
\bibitem{LH}C.X. Li, X.B. Hu, Phys. Lett. A 329 (2004) 193-198.
\bibitem{ZGH}J.X. Zhao, Math. Comput. Simul. 74
(2007) 388-396.
\bibitem{WHY}H.Y. Wang, J. Math. Anal. Appl. 330 (2007) 1128-1138.
%\bibitem{HR1}R. Hirota, Direct Method in Soliton Theory, Iwanami,
%1992 (in Japanese).
%\bibitem{HS}R. Hirota, J. Satsuma, Prog. Theor. Phys. Suppl. 59
%(1976) 64-100.
\bibitem{HR2}R. Hirota, The Direct Method in Soliton Theory (Eds.
A. Nagai, J.J.C. Nimmo and C.R. Gilson), CUP, 155, 2004.
\bibitem{MK}A.V. Mikhailov, JETP Lett. 30 (1979) 443-448.
\bibitem{NW}J.J.C. Nimmo, R. Willox, Proc. Roy. Soc. London Ser. A 453 (1997) 2497-2525.
\end{thebibliography}
\end{document}